# Recurrence plots of sunspots, solar flux and irradiance

A. Sparavigna
Dipartimento di Fisica
Politecnico di Torino
C.so Duca degli Abruzzi 24, Torino, Italy

**Abstract**

The paper shows the recurrence and cross recurrence plots of three time series, concerning data of the solar activity. The data are the sunspot number and the values of solar radio flux at 10.7 cm and of solar total irradiance, which are known as highly correlated. To compare the series, the radio flux and irradiance values are monthly averaged. Recurrence plots display the oscillating behaviour with remarkable features. Moreover, cross recurrence plots help in identifying time lags between the sunspot number maximum and the maximum of radio or irradiance signals, in circumstances where the data values are highly dispersed. Image processing is useful too, in enhancing the monitoring. An interesting behaviour is displayed by cross recurrence plots of irradiance, which are not symmetric with respect to the line of identity.

**Introduction**

Physical processes can have distinct recurrent behaviours with periodicity, or irregular and chaotic patterns. Intermittent chaotic behaviours can be eventually displayed by the dynamical system. Among the methods to analyse the system dynamics, we can find an approach based on determining the recurrence of states, which has the following meaning, that we can find states of the system that are again arbitrarily close after some time of divergence. In 1987, Eckmann et al. proposed recurrence plots, which can visualise the recurrence of states, in phase spaces with an arbitrary number of dimensions [1,2]. Eckmann's plot enables to visualise the phase space trajectory through a two-dimensional representation of its recurrences. Such recurrence of a state at time $i$, at a different time $j$, is pictured within a two-dimensional squared matrix with black and white dots, where black dots mark a recurrence. Both axes of the image are time axes. This representation is named "recurrence plot". The resulting image depends on the threshold distance, defining nearer states in the phase space.

On contrary, a recurrence plot is not depending on the scale factor, which we use to describe the phase space states. If we have to plot the recurrence of a signal, for instance a radio signal, the resulting plot will be the same for any scale factor or shift of the signal. This is the reason why we can use the recurrence to compare signals with a different observation origin. Of course, the origin must be in the same dynamical system. One example can be the comparison of El Niño Southern Oscillation and the North Atlantic Oscillation (NAO), which are observed in the climatic earth behaviour.

Because data are often non-stationary and complex, the comparison is far from simple if performed by standard time series analysis: the cross recurrence plots recently proposed can be a useful tool for solar time series [3,4]. The cross plots and image processing are what we use here to compare the behaviour of sunspot, solar radio data and irradiance. The analysis allows discussing the correlation of these sets of data.

**Sunspots and solar radio data.**

After 17 years of observations, S.H. Schwabe discovered in 1843 the cycle of the average number of sunspots. The average duration of the sunspot cycle is 11.1 years: on January 4, 2008 officially began the cycle number 24. Data on sunspots are available from the National Geophysical Data Center (NGDC), NOAA Satellite and Information Service [5]. Let us remember that a sunspot is a region on the photosphere with a lower temperature than its surrounding and an intense magnetic activity. Other data showing an oscillation are the solar radio emission and the irradiance data.

The measurement of radio waves is used to monitor the structure of the solar corona, the outer most regions of the Sun's atmosphere. Solar radio emissions at different frequencies allow observing the radiation from different heights in the solar atmosphere. A lower frequency corresponds to a higher height of origin, because the frequency decreases uniformly outwards: 245 MHz originates high in the corona, while 15,400 MHz originates in the low corona. The 5 MHz emission corresponds to about 10 solar radii height [6]. The microwave wavelength 2800 MHz daily radio flux highly correlates with the daily sunspot number. The two databases are usually considered as interchangeably. Moreover, in 1995, Schmahl and Kundu found that the solar radio fluxes in the spectral range 1000-9400 MHz correlate well with the total solar irradiance [7].

Solar Radio 10.7 cm Daily Flux (2800 MHz) noon values are obtained from the National Research Council of Canada since 1947. Until 1991 the observations were made at the Algonquin Radio Observatory, near Ottawa. Over 1990-1991 the research program was transferred to the Dominion Radio Astrophysical Observatory, near Penticton, British Columbia [8,9]. The data are then known as the Ottawa/Penticton Daily Solar Flux Values. Flux values are expressed in solar flux units (1 s.f.u. = $10^{-22}$ W/m²/Hz). The quiet sun level has a flux density, which would be observed in the absence of activity and studies suggest a quiet sun flux density of about 64 s.f.u., but rarely attained. The characteristics of the observations are surveyed in [10]. The accuracy of flux was discussed in Ref.11. The sources and emission mechanisms contributing to the 10.7 cm flux are proposed in Ref.12.

Because the data on sunspot number report monthly average values, we evaluated the same average for the 10.7 cm flux data: the upper part of Fig.1 shows the two set of data from the first of January 1955 till December 31, 2007. In the lower part of the figure, the data of flux are shifted and scaled to have the same excursion for both data sets. Let us first of all plot the recurrences of the two time series. The plots are reproduced in the Figure 2 and are clearly very close to each other. The procedure to obtain the plots from data shown in the lower part of Fig.1, is as described in Ref.3. The recurrence plot $RP$ for time-discrete variables is:

$$RP_{i,j} = \Theta\left(\varepsilon - \left\| x_i - x_j \right\|\right) \tag{1}$$

where $i,j$ are the discrete time variables, corresponding to months. $x_i$ is the variable, in this case the sunspots number or the 10.7 cm flux. The $\Theta$-function depends on the difference between the distance of two points and the predefined cut-off distance. In the case of plots in Fig. 2 and the other plots in the paper, we use 10 percent. We have the two-dimensional squared matrix with black and white dots, where black dots mark a recurrence, and both axes are time axes. The recurrence is within the defined threshold. The recurrence plot is characterised by a diagonal line, called the line of identity, simply meaning that each point is recurrent with itself.

Because the two sets of data are coming from the same dynamical system (the sun), we can try to display a cross recurrence plot, defined as:

$$CRP_{i,j} = \Theta\left(\varepsilon - \left\| x_i - y_j \right\|\right) \tag{2}$$

In *RP* the distance is between two points of the same set: in this case, that is in *CRP*, the distance is evaluate between two points of the different sets $x$ and $y$. The Fig.3 shows on the left, the cross recurrence between sunspot number and flux. The plot is rather similar to the single recurrence plots. It seems to have the line of identity too. In the cross recurrence plots, a deviation or distortion of the diagonal part of the plot from the line of identity means that the time behaviour is different [3].

Image processing can be useful too. On the right part of Fig.3, we show an image obtained combining the single recurrence plots of Fig.2: in green, the points where the two *RPs* are coincident and in red, the points where they are different. The regions *a* and *b*, corresponds to periods where we find a high solar activity. In *a* (cycle 20), the recurrences are almost equal, but in *b* (cycle 23), the red pixels evidence the difference in time when the maximum of activity is reached. The fact that 10.7 cm radio signal and sunspot number are highly correlated but not interchangeable, especially near solar maximum was reported in Ref.13. The speculative possibility proposed in Ref.13, is that the radio-flux lag discriminate between long-period and short-period cycles, an indicator of the presence of two solar cycle modes.

**Solar irradiance.**

Satellite observations demonstrated that the total solar irradiance (TSI), that is the amount of solar radiation received at the top of the earth atmosphere does vary. The TSI variations have been discussed [14]. The data we are here using are from [15]: these unpublished data are obtained in the VIRGO Experiment on the co-operative ESA/NASA Mission SOHO. Data provides the total solar irradiance in W/m$^2$ for each day from 1978 till 2003. For comparing with data of sunspot number, as in the case of 10.7 cm radio signals, we evaluate the average value on each month of the year. The Fig.4 shows in the upper part the average signal and in the lower part, the scaled and shifted data of TSI and 10.7 cm solar flux. In the Fig.5, the recurrence plot of the total solar irradiance is shown: it has the same features of the recurrences previously displayed.

The irradiance values are rather dispersed but we see that they have the same oscillating behaviour: let us compare them with the sunspot number and with the solar flux by means of the cross recurrence plots. The results of calculations are in the Fig.6. From the figure, we can see that, in spite of the rather scattered values, the two plots possess the same features. In particular, the lack of symmetry with respect to the line of identity indicates the presence of certain time lags. The author has not enough specific knowledge of solar physics to go in deep details and discuss this phenomenon: the cross recurrence plots and image processing here proposed seems to be able in identifying useful parameters for solar physics.

**Conclusions.**

We have proposed in this paper the recurrence and cross recurrence plots of three time-series data concerning the solar activity. The data of solar radio flux at 10.7 cm and of total irradiance are monthly average values, to compare with the behaviour of sunspot number. If we observed the graphs of these data, we clearly note the oscillating behaviour. The deduction from graphs of other features is not easy, due to the strong dispersion of values, which remains even after average. Recurrence plots are as fingerprints instead: according to chosen threshold, they are able to give an image of recurrence of maxima and minima. The information is the same of course, but more impressive. For instance, from the images in Fig.2, at the left side and at the bottom of each image, we note a white band telling that in cycle number 19, the maximum of activity is undoubtedly higher than that observed in other cycles. More information we find in the cross recurrence plots; with them we can compare the behaviour of two complex data sets, where the values, which belong to periods of high solar activity, are rapidly changing. In cycle number 23, we find from cross recurrence, aided with image processing, a lag in reaching the maxima. Moreover, the cross recurrences between the irradiance and the sunspot number, and between irradiance and 10.7 cm flux, reveal that plots are not symmetry with respect to the line of identity.

**Acknowledgements**

Many thanks for data collections of solar activity. We acknowledge the receipt of the dataset (composite_d25_07_0310a.dat) from PMOD/WRC, Davos, Switzerland, unpublished data from the VIRGO Experiment on the cooperative ESA/NASA Mission SOHO.

**References**

1. J.-P. Eckmann, S.O. Kamphorst, D. Ruelle, Recurrence plots of dynamical systems, Europhysics Letters 5, 973-977, 1987.

2. J.-P. Eckmann, D. Ruelle, Ergodic theory of chaos and strange attractors. Review of Modern Physics 57(3), 617-656, 1985.

3. N. Marwan, J. Kurths, Cross recurrence plots and their applications, in Mathematical Physics Research at the Cutting Edge, C.V. Benton Editor, pp.101-139, Nova Science Publishers, 2004.

4. J.P. Zbilut, C.L. Jr. Webber, Embeddings and delays as derived from quantification of recurrence plots. Phys. Letters A 171(3-4), 199-203, 1992.

5. The numerical data on sunspots are available from the NOAA NGDC Solar Data Services, at page http://www.ngdc.noaa.gov/stp/SOLAR/ftpsunspotnumber.html.

6. McLean and Labrum, Solar radiophysics: studies of emission from the sun at metre wavelengths, Cambridge University Press, 1985.

7. E.J. Schmahl, M.R. Kundu, M.R. 1995, J. Geophys. Res. 100(A10), 19851-19864, 1995; Synoptic Radio Observations, at http://www.astro.umd.edu/~ed/synoptic97/schmahl97.html.

8. The numerical data for graphs and selected bibliography are given in Algonquin Radio Observatory Report No. 5, A Working Collection of Daily 2800 MHz Solar Flux Values 1946-1976, A.E. Covington, Herzberg Institute of Astrophysics N.R.C. of Canada, Ottawa, Canada.

9. The data of the solar radio flux, noon flux measurements from Ottawa/Penticton (10.7 cm/2800 MHz) are downloaded via FTP from the NOAA NGDC Solar Data Services.

10. A.E. Covington, Solar Radio Emission at 10.7 cm, J. Royal Astron. Soc. Canada, 63, 125, 1969.

11. K.F. Tapping, D.P. Charrois, Limits to the Accuracy of the 10.7cm Flux, Solar Physics, 1994.

12. K.F. Tapping, Recent Solar Radio Astronomy at Centimeter Wavelengths: The Temporal Variability of the 10.7-cm Flux, J. Geophys. Res., 92(D1), 829-838, 1987.

13. R.M. Wilson, D. Rabin, R.L. Moore, 10.7 cm solar radio flux and magnetic complexity of active regions, Solar Physics, 111(2), 279-285, 2004

14. C. Frölich and J.Lean, The sun total irradiance: cycles, trends and related climate change uncertainties since 1978, Geophys. Res. Lett. 25, 4377-4380, 1998

15. Data are downloaded at FTP/STP/SOLAR_DATA/SOLAR_IRRADIANCE in ftp.ngdcnoaa.gov

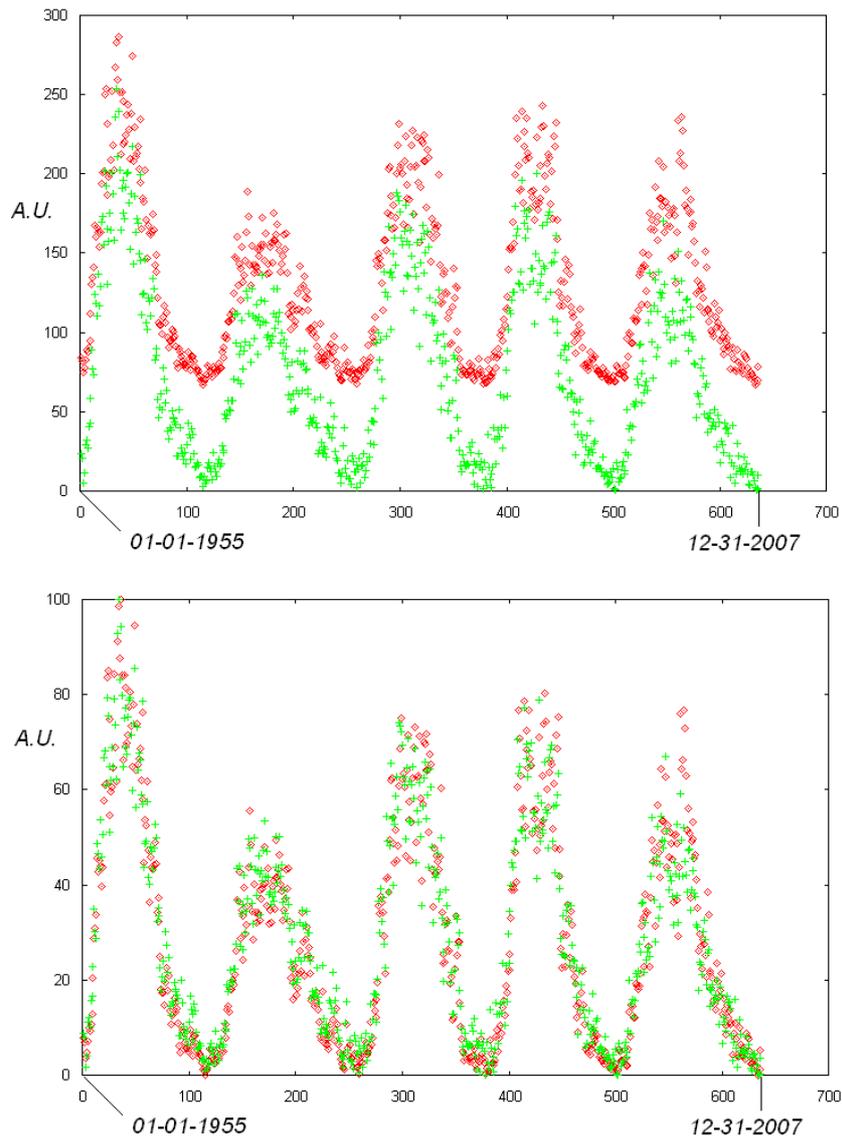

Fig.1 Monthly average values of sunspot number (red) and 10.7 cm flux data (green). The first image shows the two sets of data from the first of January 1955 till December 31, 2007. In the lower part of the figure, the data of flux are shifted and scaled.

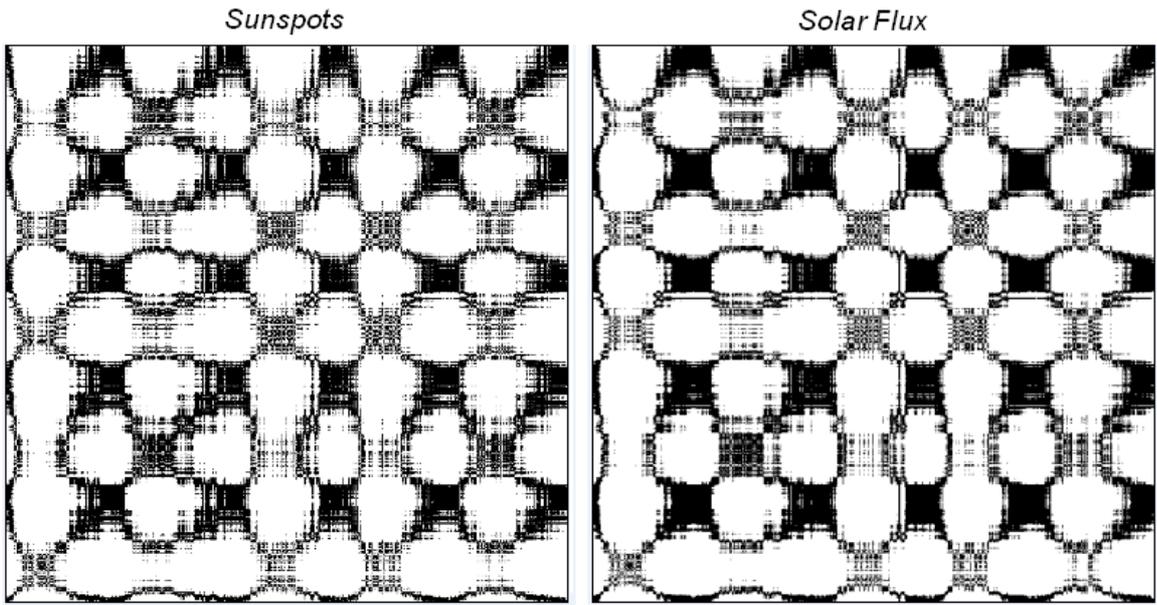

Fig. 2 The image shows the two-dimensional squared matrix with black and white dots, where black dots mark a recurrence for data of sunspot numbers (on the left) and solar flux (on the right). Both axes are time axes. The recurrence is within a threshold of 10 percent. Note the almost equal behaviour. Data are from the first of January 1955 till December 31, 2007.

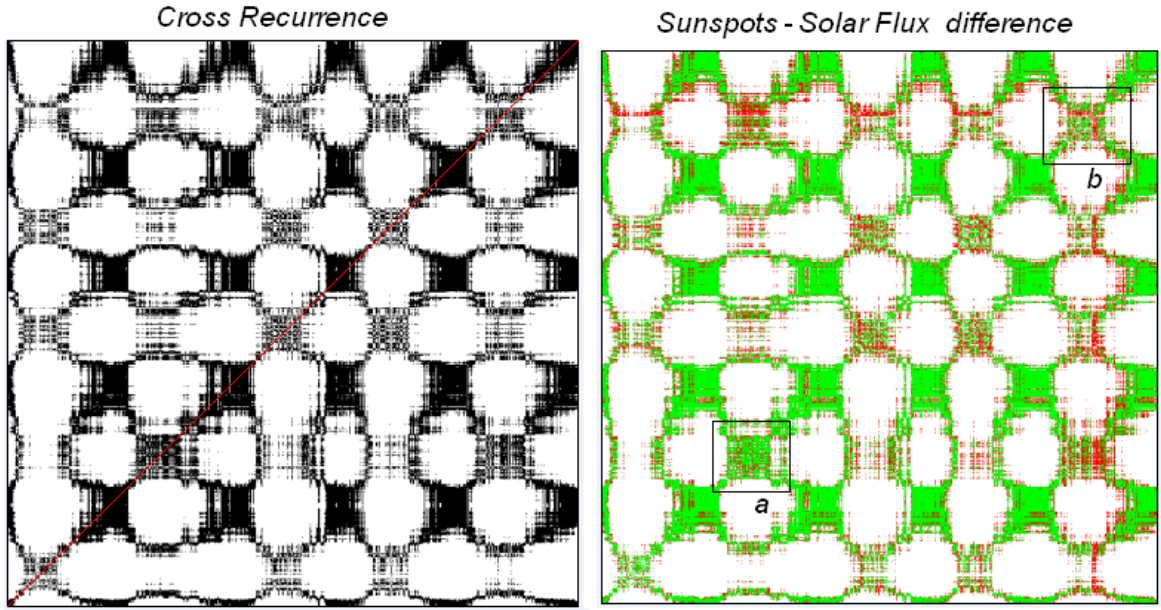

Fig. 3 On the left, the image shows *CRP* of sunspots and solar flux. The recurrence is within a threshold of 10 percent. The image looks like the single recurrence plots, but differences are observable (in red, the line of identity). On the right, we show an image obtained combining the single recurrence plots of Fig.2: in green, the points where the *RPs* are coincident and in red, the points where they are different. The regions *a* (cycle 20) and *b* (cycle 23) correspond to two maxima of the solar activity. In *a*, the recurrences are almost equal, but in *b*, the red pixels evidence a difference in time lag.

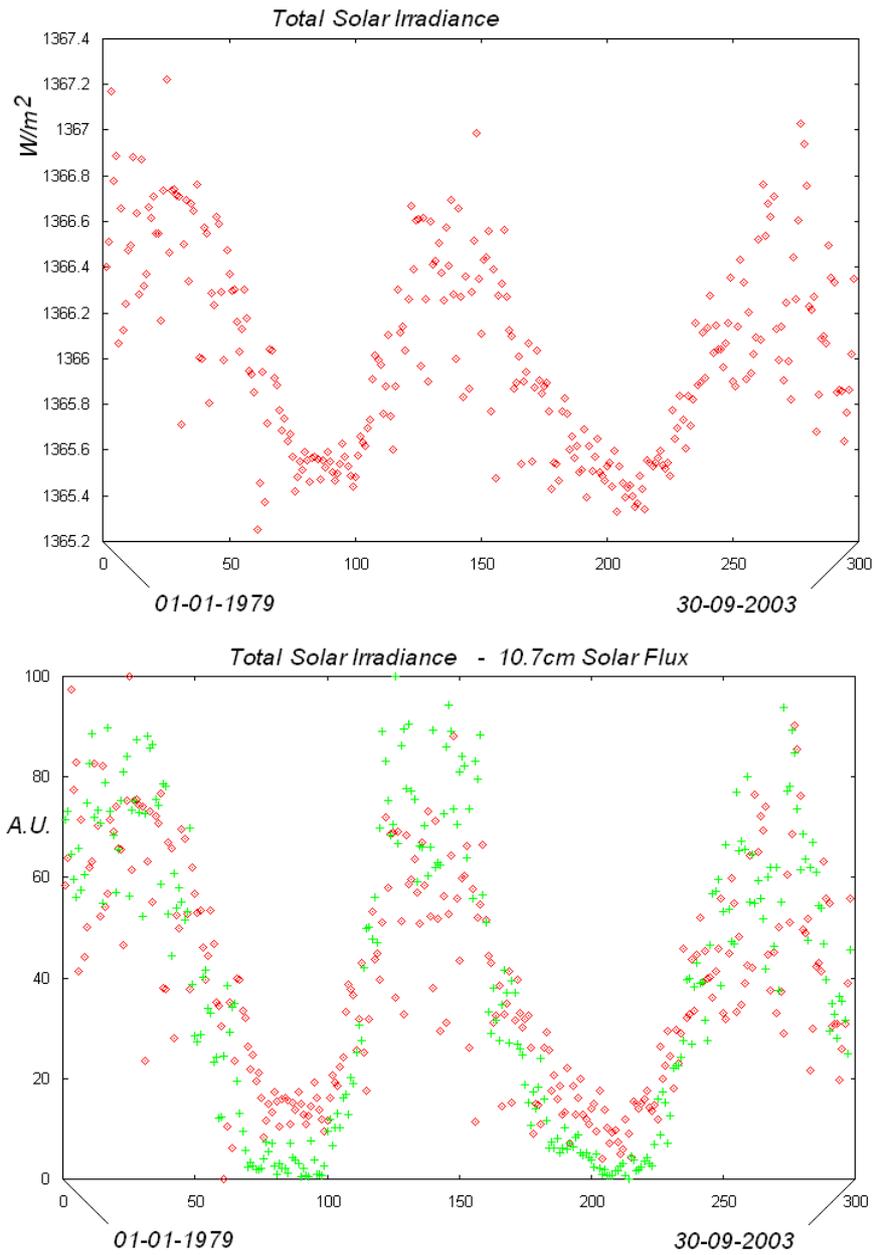

Fig.4 Monthly average values of Solar Total Irradiance in the upper part, as from data [15]; in the lower part the irradiance (red) and 10.7 cm flux data (green), scaled and shifted. Data are from the first of January 1979 till September 30, 2003.

## Recurrence Irradiance

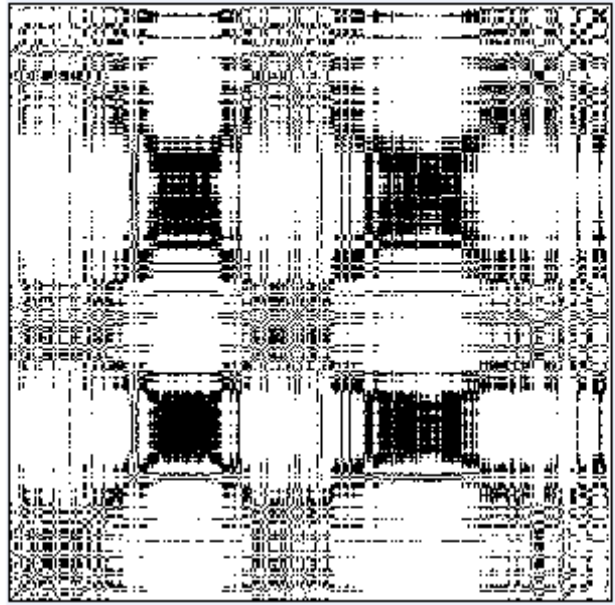

Fig.5 The recurrence plot of the Solar Total Irradiance (from data [15]). Data are from the first of January 1979 till September 30, 2003. The darker areas are corresponding to minima of solar activity.

## Cross Recurrence

### Sunspots - Irradiance

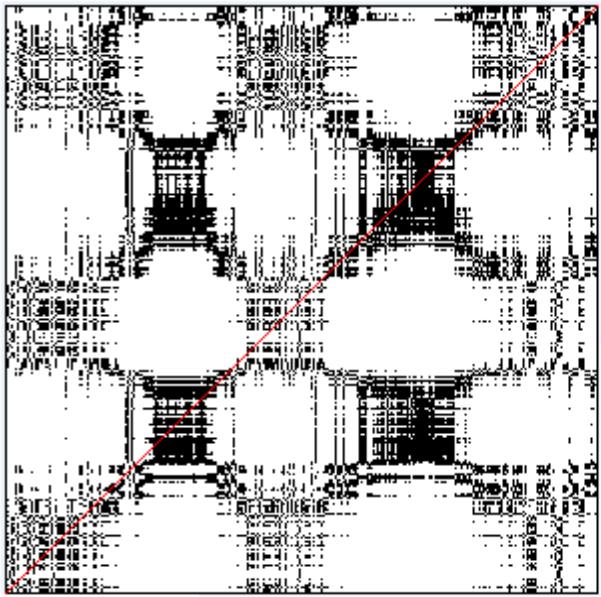

### 10.7cm Solar Flux - Irradiance

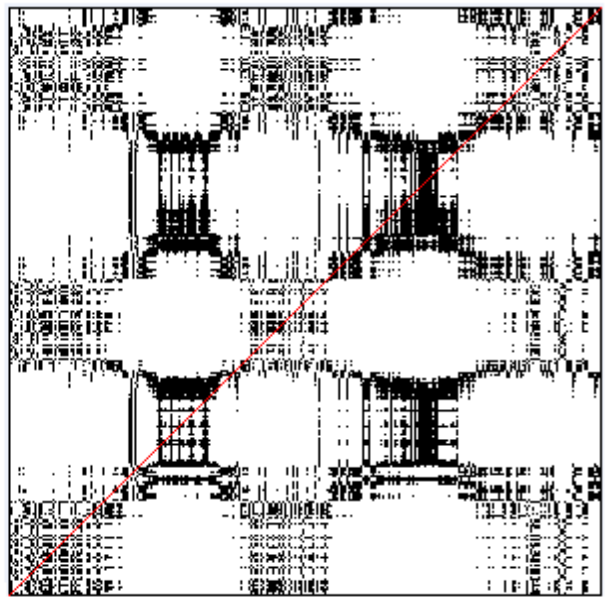

Fig.6 The cross recurrence plots of the Solar Total Irradiance with the sunspot number oscillation and the 10.7 cm signal. Data are from the first of January 1979 till September 30, 2003. Note that, in spite of the rather scattered data of irradiance, the two plots have the same features. Note that the plots are not symmetric with respect to the line of identity.